# Directionality Fields generated by a Local Hilbert Transform

W. W. Ahmed[1,2], R. Herrero[1], M. Botey[1], Z. Hayran[3], H. Kurt[3] and K. Staliunas[1,4]

[1]*Departament de Física, Universitat Politècnica de Catalunya (UPC), Colom 11, E-08222 Terrassa, Barcelona, Spain*
[2]*European Laboratory for Non-linear Spectroscopy (LENS), Sesto Fiorentino 50019, Florence, Italy*
[3]*Department of Electrical and Electronics Engineering, TOBB University of Economics and Technology, Ankara 06560, Turkey*
[4]*Institució Catalana de Recerca i Estudis Avançats (ICREA), Passeig Lluís Companys 23, E-08010, Barcelona, Spain*

We propose a new approach based on a local Hilbert transform to design non-Hermitian potentials generating arbitrary vector fields of directionality, $\vec{p}(\vec{r})$, with desired shapes and topologies. We derive a local Hilbert transform to systematically build such potentials, by modifying background potentials (being either regular or random, extended or localized). In particular, we explore particular directionality fields, for instance in the form of a focus to create sinks for probe fields (which could help to increase absorption at the sink), or to generate vortices in the probe fields. Physically, the proposed directionality fields provide a flexible new mechanism for dynamically shaping and precise control over probe fields leading to novel effects in wave dynamics.

Systems described by non-Hermitian potentials, first introduced in quantum mechanics and linear electrodynamics [1,2], have recently found realizations in optics [3-8], by using coherent gain and losses, thus opening the new discipline of non-Hermitian optics. One of the most fascinating properties of such non-Hermitian systems is that the parity (in one dimension), or generally the space symmetry (in two or higher dimensions) can be broken at around so named exceptional points. As a consequence, different counterintuitive physical effects arise, such as unidirectional invisibility [6-8], unidirectional transmission [9], unidirectional lasing [10,11] antibandgaps [12], perfect unidirectional absorption [13,14], nonreciprocal Bloch oscillations [15], and generally the unidirectional transfer of energy in linear [3-5] and nonlinear [16-18] systems. Most of the new intriguing features were initially proposed in a particular kind of such non-Hermitian systems, namely in those holding PT-symmetry [1].

An interesting extension of such broken-symmetry systems is when such unidirectionality holds only locally [19]. The simplest case of such one-dimensional (1D) system is, for instance, when the locally broken symmetry favors wave propagation to the left on the right half-space, and to the right on the left half-space. Such a potential may be expected to introduce a sink at the boundary of both domains with different parity, i.e. at *x*=0, where the probe field may be efficiently localized [19]. Field localization due to favored flows towards a selected position, see Fig.1(a), may be extremely desirable either in linear and nonlinear optics, especially beyond 1D, in higher-dimensional systems. We propose here general locally non-Hermitian potentials, see Fig. 1(b), where the probe fields could be configured into arbitrary shapes and flows, for instance into a sink, a vortex, or into closed flow channels, see Fig.1(c,d). Such potentials hold local PT-symmetry in the entire spatial domain. Therefore, working exactly on the local exceptional point can allow a precise control over the field flow sense, enabling the generation of arbitrary directionality fields. The general ultimate goal is to systematically construct arbitrary complex non-Hermitian potentials favoring any desired configuration of directionality fields in the system.

This letter provides a solution of the problem sketched above. We first consider a given directionality field $\vec{p}(\vec{r})$, as denoted by the arrows in Fig.1. Next, we consider any initial background potential $n(\vec{r})$, being, for instance, real (in the case of optics, a real refraction index profile), or being, more generally, also complex (by including optical gain and losses). Such an initial background potential may correspond to a localized object on a constant refraction index background, $n(\vec{r}) \to n_0$ for $|\vec{r}| \to \infty$, as well as to spatially extended patterns, i.e. a periodic, quasiperiodic or randomly distributed background potential. The central message of the letter is to propose and derive the explicit integral relation, which could be referred as a local Hilbert transform, to obtain the reciprocal part of the background potential, $m(\vec{r})$, which ensures a desired configuration of the directionality field, defined by any arbitrary vector field, $\vec{p}(\vec{r})$. For instance, in the case of optics, starting from an initial refraction index profile $n(\vec{r})$, the proposed local Hilbert transform generates the corresponding spatial profile of the gain-loss function, $m(\vec{r})$, see Fig. 1(a). The local directionality of the final complex potential is warranted by the corresponding Fourier transform (zero in one half-space) as indicated in Fig.1. Finally, in order to verify the expected effect, we numerically check that the flows of the probe fields follow the given directionality field $\vec{p}(\vec{r})$, as for instance the field accumulates at around the focal points of $\vec{p}(\vec{r})$. These tests have been performed by numerically solving the Schrödinger equation (for paraxial optics, or zero temperature Bose condensates) for linear systems or the Complex Ginzburg-Landau equation (CGLE) for nonlinear systems (for instance, for broad aperture lasers) with given complex potentials. Moreover (see Supplemental Material 2 [28]), the tests hold for more complex models for tests flow, like the set of Maxwell equations.

To characterize the complex directionality fields, we introduce a measure of local directionality. In the simplest 1D case, a scalar function $p(x)$ is sufficient, equal to $+1/-1$ for right/left directionality, respectively. Such uniform directionality ($p(x) = const = \pm 1$) can be constructed by a pair of the well-known Hilbert transform:

$$m(x) = \frac{-p}{\pi} \mathcal{P} \int \frac{n(x_1)}{x-x_1} dx_1$$

$$n(x) = \frac{-p}{\pi} \mathcal{P} \int \frac{m(x_1)}{x-x_1} dx_1, \quad (1)$$

connecting the two quadratures of a complex function in physical space. The quadratures can be, for instance, the real and imaginary parts of the potential (corresponding to the refraction index and gain-loss in optics). The symbol $\mathcal{P}$ means the principal value of the integral (the integral is calculated everywhere, except in at the singularity point).

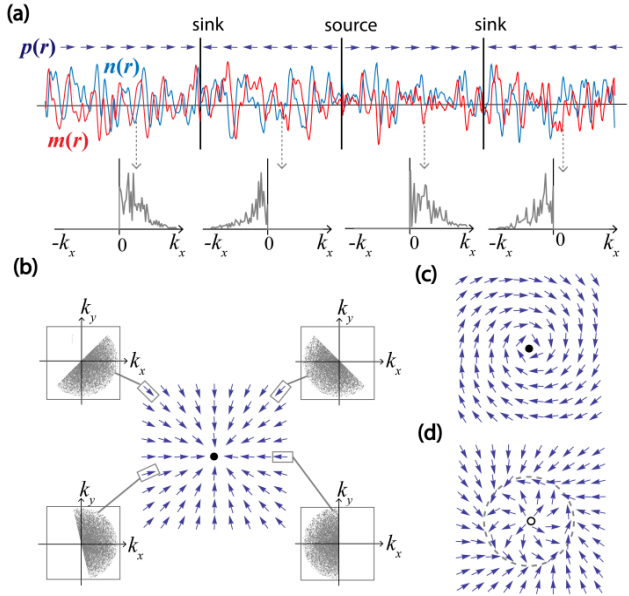

FIG. 1. (color online) Directionality fields. (a) One-dimensional directionality field consisting of spatial domains of different parity, containing sinks, and sources between the domains. The first row represents a random complex potential with given real, $n(x)$ and imaginary part, $m(x)$ (blue, and red, respectively) with corresponding Fourier spectra of each domain shown in the second row. (b) Vector field to generate a complex directionality field in two spatial dimensions in form of a focus, which may eventually create a sink for the probe field. The insets schematically represent the Fourier transforms of an arbitrary random directionality potential at several points. (c,d) Vector fields in form of a node and an antinode, resulting in vortex with sink and a circular flow channel for the probe field. In all cases, the vector fields, $\vec{p}(\vec{r})$, are denoted by the dark blue arrows.

In fact, the Hilbert transform in frequency domain (referred as the Kramers-Kronig relation) is directly related to the directionality of time. Indeed, the causality principle entails total invisibility of the future, which generates analogous relations in frequency. Analogously, the spatial Hilbert transform is related with the unidirectional invisibility in space [20,21]. In the particular case of Eq. (1), it ensures the suppression of either the left- or right-scattering by the corresponding potential, depending on the sign of $p$.

Now, in pursuit of a potential that determines an arbitrary directionality of the field, we propose a local Hilbert transform, where the $p(x)$ function can vary in space. For this general 1D case, the pair of Hilbert transforms (see Supplemental 1.1 for the derivation [22]) reads:

$$m(x) = \frac{1}{\pi} \mathcal{P} \int \frac{p(x)n(x_1)}{x-x_1} dx_1;$$

$$n(x) = \frac{-1}{\pi} \mathcal{P} \int \frac{p(x_1)m(x_1)}{x-x_1} dx_1. \quad (2)$$

Note that the directionality function enters into Eq. (2) asymmetrically, which warrants that the sequentially applied direct- and inverse local Hilbert transform recovers the initial function, while $p(x)^2 = 1$. Symbolically, Eq. (2) (see Supplementary 1.2) can be cast in the operator form: $m(x) \equiv \widehat{H}\widehat{P}n(x)$, and $n(x) \equiv \widehat{P}\widehat{H}^{-1}m(x)$ respectively. We note that generally the local Hilbert operator $\widehat{H}$ and directionality operator $\widehat{P}$ do not commute. The noncommutivity of these operators require the asymmetric form of transformations, Eq. (2). This noncommutivity problem does not occur in classical Hilbert transform pair for uniform directionality.

Finally, we tackle 2D systems. We define a unit directionality vector field $\vec{p}(\vec{r})$, such that $|\vec{p}(\vec{r})| = 1$. Then, the pair of local Hilbert transforms reads (see Supplemental 1.3 [22]):

$$m(\vec{r}) = \frac{1}{\pi} \mathcal{P} \iint \frac{\delta((\vec{r}-\vec{r}_1)\cdot\vec{q}(\vec{r}-\vec{r}_1))n(\vec{r}_1)}{\vec{p}(\vec{r})(\vec{r}-\vec{r}_1)} d\vec{r}_1 \, ;$$

$$n(\vec{r}) = \frac{-1}{\pi} \mathcal{P} \iint \frac{\delta((\vec{r}-\vec{r}_1)\cdot\vec{q}(\vec{r}-\vec{r}_1))m(\vec{r}_1)}{\vec{p}(\vec{r}_1)(\vec{r}-\vec{r}_1)} d\vec{r}_1. \quad (3)$$

Here $\vec{q}(\vec{r})$ is a unit vectorial field orthogonal to a given field of directionality: $\vec{q}(\vec{r}) \cdot \vec{p}(\vec{r}) = 0$, and $|\vec{q}(\vec{r})| = 1$. Note again that $\vec{p}(\vec{r})$ and $\vec{p}(\vec{r}_1)$ enter asymmetrically into Eq. (3). Besides, technically, due to the presence of the Kronecker-delta function in Eq. (3), the integrals become, effectively, one-dimensional.

The pair of local Hilbert transforms, Eq. (3), is our central result, which is used throughout the rest of the letter to design specific potentials to generate the given fields of directionality, and to explore the propagation of the probe fields in such potentials.

Although Eq. (3) may relate arbitrary (orthogonal) quadratures of the complex potential, we here consider a simple case where the quadratures are the real and imaginary parts of the potential: $n(\vec{r}) \equiv n_{re}(\vec{r})$, $m(\vec{r}) \equiv n_{im}(\vec{r})$, i.e. the refraction index and gain-loss spatial distributions of the corresponding optical potentials. We start from a given refraction index profile $n_{re}(\vec{r})$ (either describing localized objects, extended objects or a random background), and then build the corresponding gain-loss profile, $n_{im}(\vec{r})$, by using equation (3) to obtain a final complex potential $n(\vec{r}) = n_{re}(\vec{r}) + in_{im}(\vec{r})$, which ensures a given directionality field $\vec{p}(\vec{r})$. Then, we probe the potential by numerically solving the Schrodinger equation with corresponding complex potential (see Supplemental Material 2 [28] for the other models of probing field).

We consider a paraxial electromagnetic field equation of diffraction (equivalent to the Schrödinger equation for a quantum wave-function) for an optical beam including the non-Hermitian potential, $U(\vec{r})$ in the normalized form as:

$$\partial_t A(\vec{r},t) = i\nabla_\perp^2 A(\vec{r},t) + iU(\vec{r})A(\vec{r},t) \quad (4)$$

where $A(\vec{r},t)$ is the slowly varying electric field envelope distributed in, $\vec{r}$, and evolving in time, $t$ and $U(\vec{r}) = n_{re}(\vec{r}) + in_{im}(\vec{r})$ is a complex potential in two-dimensional space generated from arbitrary vector field of directionality, $\vec{p}(\vec{r})$. Here, $n_{re}(\vec{r})$ is the initial considered background pattern and $n_{im}(\vec{r})$ is the constructed imaginary part of the potential using proposed local Hilbert transform. We numerically solve Eq.(4) using split step method to explore different directionally fields that demonstrate the functionality of our proposal.

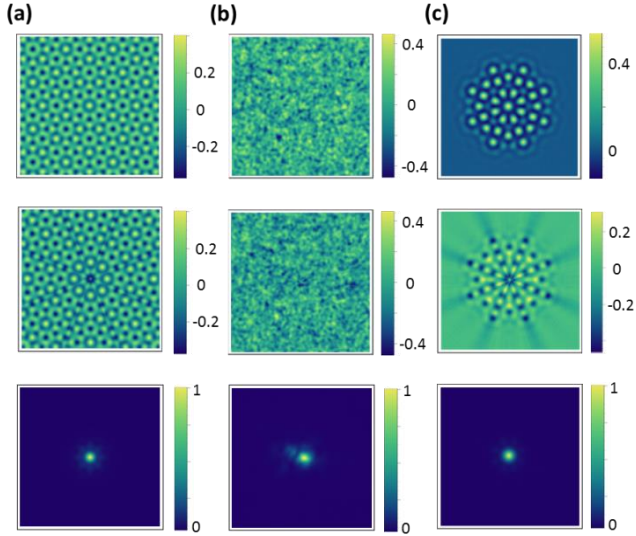

FIG. 2. (color online) Probe field evolution in directionality fields. Directionality fields in form of a focus (the one shown in Fig. 1(b)) generated from the real-valued background potential with different profiles such as: (a) an octagonal pattern, (b) a random pattern, (c) a localized pattern. The upper row depicts the real part of the background potential, $n_{re}(\vec{r})$, the second row the corresponding imaginary part of the potential generated by the local Hilbert transform, $n_{im}(\vec{r})$, and the bottom row shows the normalized final distributions of the probe field after sufficiently long time. The evolutions of the field for the three cases is shown in the corresponding movie (Supplemental Movie 1 [23])

We first consider the most simple directionality field in form of a sink or a focus (see Fig. 1(b)): $\vec{p}(\vec{r}) = -\vec{r}/|\vec{r}|$, starting from different background potentials, $n_{re}(\vec{r})$. The numerical results for an octagonal, random and localized background potential are presented in Fig. 2. For a particular initial probe field (a Gaussian beam at any arbitrary position) we observe the flow of the probe field towards the center, $\vec{r} = 0$. After a sufficient evolution time, the field is finally localized at the center in all cases, yet it grows in time due to linearity of the system while maintaining its spatial profile. We note the robustness of the mechanism, since localization occurs irrespectively from the initially chosen probe field position. Also different background configurations (octagonal, random, localized) lead to similar localization distributions, with small differences on small space scales due to different symmetries of the background potential.

Next, we explore more exotic directionality fields: a superposition of vortex with a sink as defined by the vector field: $\vec{p}(\vec{r}) = -(\alpha\vec{r} + \beta\vec{r} \times \vec{e}_z)/|\alpha\vec{r} + \beta\vec{r} \times \vec{e}_z|$, and a circular channel flow as: $\vec{p}(\vec{r}) = (\alpha(|\vec{r}|-1)\vec{e}_r + \beta\vec{r} \times \vec{e}_z)/|\alpha(|\vec{r}|-1)\vec{e}_r + \beta\vec{r} \times \vec{e}_z|$. In both expressions $\vec{e}_z$, $\vec{e}_r$, and $\vec{e}_\varphi$ stand for the unit basis vectors in polar coordinates, and the $\alpha$ and $\beta$ correspond to the radial and azimuthal part of the flow, respectively. The results for both cases are provided in Fig. 3.

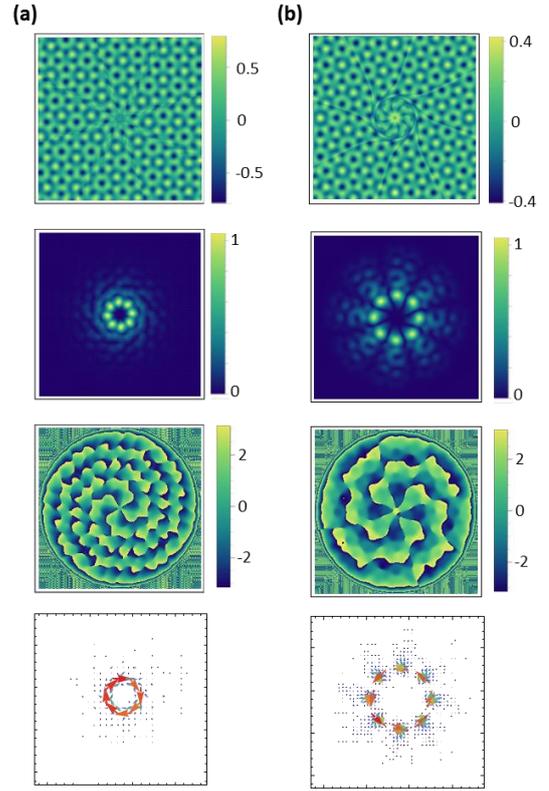

FIG. 3. (color online) Probe field evolution in chiral directionality fields for background octagonal potential. (a) vortex with sink (b) circular flow channel, corresponding to the vector fields of Fig. 1(c) and 1(d). The first row represents the imaginary part of the potential (real part is shown in Fig.2; the second row shows the final amplitude distributions of the probe field, the third row shows the phase distributions of the probe field and the fourth row represents the corresponding flows of the probe field. The field flow is determined by: $-i(A^*\nabla A - A\nabla A^*)$, where $A(\vec{r},t)$ is the spatial distribution of the probe field.

The final spatial field distributions in Fig. 3 indicate that the probe field is accumulated at the center, $\vec{r} = 0$, in case of a vortex-sink in directionality field and on a ring for the directionality field in form of a circular channel. The evolutions of the field towards the final state in the considered chiral cases is also robust (weakly dependent on the background potential as well as on initial distribution of probe fields), and is shown in the corresponding movie (Supplemental Movie 2 [23]).

Next, as a proof of concept and to show the applicability of the proposed scheme also for nonlinear systems (not only for linear systems, as above demonstrated for the Schrödinger equation), we study a possible implementation for laser-like systems, with field saturation. We consider a patterned Vertical Cavity Surface Emitting Laser (VCSEL), as shown in Fig. 4(a), where the sink functionality is implemented to improve the brightness output beam. The system may be described by the CGLE with the corresponding complex potential [19]. The results show the field localization induced by the unidirectional potential, see Fig. 4(b). It is clear from the flux depicted in the inset of Fig. 4(b), that field is directed towards the center, confirming the sink behavior, leading to a bright and stable output beam when the field is saturated.

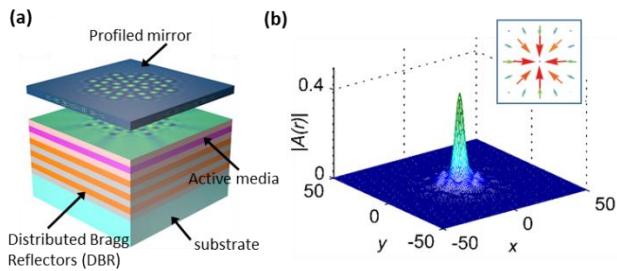

FIG. 4. (color online) Field evolution of the VCESL described by the CGLE (the laser case). (a) Schematic of a modified VCSELs with a patterned active layer and a correspondingly profiled mirror may lead to a directionality field in the form of a localized pattern with sink, as shown in Fig. 2(c). (b) Final distribution of the output field after saturation for $p$=-0.1. The inset represent the field flux for the corresponding field distribution. Note that the spatial domain ($x,y$) have normalized units.

To conclude, we propose and derive a local Hilbert transform relating the two quadratures of the non-Hermitian potential, in order to construct the arbitrary directionality fields from any arbitrary background pattern. The results obtained in 2D, could be straightforwardly extendable to 3D systems. We present the simplest cases of such directionality fields in form of a sink, a vortex, and a circular flow channel, while, evidently, the procedure allows building directionality fields of any shape. We also show that the local Hilbert transform can be applied on initial periodic, random and localized background potentials to generate such directionality fields, among others. The proposed theory can yield a significant control over the electromagnetic field flows, providing an alternative to the broad field of transformation optics [24-27]. Our proposal is generic in nature, which may offer rich possibilities for structuring the light in linear and nonlinear micro-optical systems, and could be applied to all fields of wave dynamics, like zero temperature Bose condensates, acoustics, plasmonics waves, etc., all described by the models of Schrödinger equation type.

We note that the potentials ensure the directionality fields not only probing by Schrödinger equation, but also by other models, for instance the full Maxwell equations governing electromagnetic fields without paraxial approximation (see Supplemental Material 2 [28]). The specific model alters only the Green function in scattering theory, however does not affect the scattering potentials and, consequently the local Hilbert transform is unaffected.

Note also that an analogous directionality field may be achieved with only purely lossy media, thus increasing the feasibility for practical realization e.g; Perfectly Matched Layer (PML) on arbitrary boundary contours. Moreover, directionality fields can be generated in nonlinear cases. For instance we demonstrate that the directionality fields in form of a sink may increase the brightness of the emission of broad aperture lasers, leading to bright and narrow output beams. All this indicates a huge application potential, apart from a fundamental importance of the local Hilbert transform itself.

Authors acknowledge financial support of NATO SPS research grant No: 985048, support from Spanish Ministerio de Ciencia e Innovación, and European Union FEDER through project FIS2015-65998-C2-1-P, Erasmus Mundus Doctorate Program Europhotonics (Grant No. 159224-1-2009- 1-FR-ERA MUNDUS-EMJD) and partial support of the Turkish Academy of Sciences.

# Supplemental Material 1

In this supplementary material, we derive the local Hilbert transform for two-dimensional (2D) potentials, systematically exploring the cases in increasing order of complexity, starting from the well-known Hilbert transform in 1D case, to the local Hilbert transform in 1D, then finally extending into 2D case.

## 1.1. Spatial Hilbert transform in 1D

Let us consider the background potential, $n(x)$, with its spatial Fourier transform $n(k_x) = \frac{1}{\sqrt{2\pi}} \int n(x) exp(-ik_x x) dx$. The space symmetry is maximally broken if the spectra is set to zero either on the left- or right half-axis $k_x$, depending on the value of $p = \pm 1$. This can be done by adding to the initial potential a potential with the following spectrum:

$$m(k_x) = n(k_x) \cdot p \cdot sgn(k_x) \tag{S1}$$

where $sgn(*)$ is the sign-function. The antisymmetric part of the spectrum (with the sign function) corresponds to the imaginary part of the potential in the real space, if the initial potential $n(k_x)$ is a real one. (S1), the product of two functions, is equivalent to the Hilbert transform in wavenumber $k_x$ domain. The standard form of the Hilbert transform in x-domain is the convolution of the inverse Fourier spectra of both product components of (S1):

$$m(x) = \frac{p}{\pi} \mathcal{P} \int \frac{n(x_1)}{x-x_1} dx_1 \tag{S2}$$

Note that the inverse Fourier transform of the sign-function is $\sqrt{2/\pi} \cdot i/x$, and the convolution in (S2) in 1D introduces the factor $(2\pi)^{-1/2}$; $\mathcal{P}$ means the principal value of the integral. Differently from the conventional form of the Hilbert transform we introduced here the directionality factor $p$. Analogously, the inverse Hilbert transform reads:

$$n(x) = \frac{-p}{\pi} \mathcal{P} \int \frac{m(x_1)}{x-x_1} dx_1 \tag{S3}$$

For consistency, the sequentially direct- and inverse Hilbert transform $H * H^{-1}$ can be calculated applying sequentially (S2) and (S3):

$$m(x) = \frac{p}{\pi} \mathcal{P} \int \frac{n(x_1)}{x-x_1} dx_1 = \frac{-p^2}{\pi^2} \mathcal{P} \int \frac{m(x_2)}{(x-x_1)(x_1-x_2)} dx_1 dx_2 \tag{S4}$$

Since the relation $\frac{1}{\pi^2} \mathcal{P} \int \frac{dx_1}{(x-x_1)(x_1-x_2)} = -\delta(x-x_2)$ holds, the sequentially direct- and inverse Hilbert transform results in an identity transform.

The Hilbert transform ensures the maximum spatial asymmetry in the systems response to the non-Hermitian potentials. Hilbert transform is valid for the PT-symmetric potentials at the so-called PT-phase transition (or critical) point (but not vice versa).

## 1.2. Spatial local Hilbert transform in 1D

In case of potentials with local directionality, for instance for those containing domains of alternating directionality as shown in Fig.1.a, one should introduce the local function of the directionality, therefore the parameter $p$ becomes space-dependent $p(x)$. Then, in an analogous way, we define the local Hilbert transform:

$$m(x) = \frac{1}{\pi} \mathcal{P} \int \frac{p(x) \cdot n(x_1)}{x-x_1} dx_1 \tag{S5}$$

and the inverse:

$$n(x) = \frac{-1}{\pi} \mathcal{P} \int \frac{p(x_1) \cdot m(x_1)}{x-x_1} dx_1 \tag{S6}$$

Note that the directionality function $p(x)$ enters into (S4) and in (S5) in different ways: in (S5) it enters after the integration, and in (S6) – before the integration. The symmetric shape of the pair of Hilbert transform would result in the following integral on $\int \frac{p(x_1)dx_1}{(x-x_1)(x_1-x_2)}$ checking the sequentially direct- and inverse Hilbert transform. The above integral does not lead to $\delta(x - x_2)$ function for noncontinuos $p(x_1)$. Note, that $p(x_1)$ is not continuous at the sink and source points (see Fig.1 of the main manuscript).

Equations (S5) and (S6) written in symbolical operator form read: $m(x) = \widehat{H}\widehat{P}n(x)$, and $n(x) = \widehat{P}\widehat{H}^{-1}m(x)$, where $\widehat{H}$ is the Hilbert operator (i.e. setting of half-spectra to zero) and $\widehat{P}$ is the directionality operator (saying which half of spectra to set to zero). Calculation of the to sequentially direct- and inverse Hilbert transform result either in $m(x) = \widehat{H}\widehat{P}\widehat{P}\widehat{H}^{-1}m(x)$, or in $n(x) = \widehat{P}\widehat{H}^{-1}\widehat{H}\widehat{P}n(x)$. Since $p(x)^2 = 1$, then both direct- and inverse Hilbert transforms result in unity relations. In contrary, the direct- and inverse Hilbert transform in symmetrical case would lead to operators $\widehat{H}\widehat{P}\widehat{H}^{-1}\widehat{P}$ and $\widehat{H}^{-1}\widehat{P}\widehat{H}\widehat{P}$ which, in general, are not unity operators if $\widehat{H}$ and $\widehat{P}$ do not commute.

## 1.3. Spatial Hilbert transform in 2D

In two-dimensional systems, we define the directionality by a vector $\vec{p}$, with the norm $|\vec{p}| = 1$. The background potential is also defined in 2D space $n(\vec{r})$, and its Fourier image: $n(\vec{k}) = \frac{1}{2\pi} \int n(\vec{r}) exp(-i\vec{k}\vec{r}) d\vec{r}$. Then the directionality can be achieved by setting a corresponding half-plane of spectrum $n(\vec{k})$ to zero, which mathematically reads:

$$m(\vec{k}) = n(\vec{k}) \cdot sgn(\vec{k} \cdot \vec{p}) \qquad (S7)$$

The Fourier transform of $sgn(\vec{k} \cdot \vec{p})$ reads: $\frac{2i\delta(\vec{r}\cdot\vec{q})}{\vec{r}\cdot\vec{p}}$, where $\vec{q}$ is the unit vector orthogonal to the given field of directionality: $\vec{q} \cdot \vec{p} = 0$, and $|\vec{q}| = 1$. The Hilbert transform then follows directly:

$$m(\vec{r}) = \frac{1}{\pi}\mathcal{P} \iint \frac{\delta((\vec{r}-\vec{r}_1)\cdot\vec{q})n(\vec{r}_1)}{(\vec{r}-\vec{r}_1)\cdot\vec{p}} d\vec{r}_1; \qquad (S8)$$

$$n(\vec{r}) = \frac{-1}{\pi}\mathcal{P} \iint \frac{\delta((\vec{r}-\vec{r}_1)\cdot\vec{q})m(\vec{r}_1)}{(\vec{r}-\vec{r}_1)\cdot\vec{p}} d\vec{r}_1 \qquad (S9)$$

Equations (S8), (S9) correspond to the Hilbert transform for a uniform directionality. In order to write the final result, the local Hilbert transform in 2D, we proceed as in the 1D case, i.e. we assume that the fields $\vec{p}(\vec{r})$ and $\vec{q}(\vec{r})$ are functions of space. Then, the corresponding formulas read:

$$m(\vec{r}) = \frac{1}{\pi}\mathcal{P} \iint \frac{\delta((\vec{r}-\vec{r}_1)\cdot\vec{q}(\vec{r}-\vec{r}_1))n(\vec{r}_1)}{\vec{p}(\vec{r})(\vec{r}-\vec{r}_1)} d\vec{r}_1; \qquad (S10)$$

$$n(\vec{r}) = \frac{-1}{\pi}\mathcal{P} \iint \frac{\delta((\vec{r}-\vec{r}_1)\cdot\vec{q}(\vec{r}_1))m(\vec{r}_1)}{\vec{p}(\vec{r}_1)(\vec{r}-\vec{r}_1)} d\vec{r}_1 \qquad (S11)$$

Here $\vec{q}(\vec{r})$ is the field orthogonal to the given directionality field: $\vec{q}(\vec{r}) \cdot \vec{p}(\vec{r}) = 0$, and $|\vec{q}(\vec{r})| = 1$.

Analogously to 1D case previously explored, the sequence of direct- and inverse Hilbert transform using asymmetric pair (S10), (S11) leads to the unity relation, and thus ensures compatibility.

Note that, technically, due to the presence of the Kronecker-delta function in the integrals, they become effectively one-dimensional integrals.

# Supplemental Material 2

The purpose of this supplementary material is a numerical evidence of the accuracy of the proposed generalized local Hilbert transform beyond the paraxial approximation, therefore using the full set of Maxwell's equations for the evolution of the probe field. First, we discuss the similarities and differences between the scattering of electromagnetic fields in Helmholtz (Maxwell equations) and Schrödinger models. Next, we provide the observations of the directionality fields at optical frequencies by means of full-wave simulations employing the finite-difference time-domain (FDTD) method, confirming that our theory works also on Maxwell equations. Besides, we finally show that such a wave operation can also be realized via a pure lossy medium, thereby increasing the feasibility and variety of potential practical realizations.

## 2.1. Scattering properties of Helmholtz and Schrödinger equations

We consider a scalar Helmholtz equation, as directly derived from Maxwell equations:

$$\nabla^2 E(\vec{r}) + \varepsilon(\vec{r}) k_0^2 E(\vec{r}) = 0 \tag{S1}$$

where $k_0 = \omega_0/c$ is the wave number of frequency $\omega_0$ and $\varepsilon(\vec{r}) = \varepsilon_0 + \Delta\varepsilon(\vec{r})$ is a weakly modulated dielectric permittivity i.e $\Delta\varepsilon(r) \ll \varepsilon_0$, where $\varepsilon_0$ is the positive background (relative) dielectric permittivity. When considering a complex permittivity, $\Delta\varepsilon = \Delta\varepsilon_{Re} + i\varepsilon_{Im}$, the system becomes non-Hermitian. Then, eq. (S1) becomes:

$$\nabla^2 E(\vec{r}) + \varepsilon_0 k_0^2 E(\vec{r}) = -k_0^2 \Delta\varepsilon(r) E(\vec{r}) \tag{S2}$$

Performing the Fourier transform of both sides, we obtain the following implicit equation for the scattered field:

$$E_s(\vec{k}) = \frac{k_0^2}{(k^2 - \varepsilon_0 k_0^2)} \frac{1}{2\pi} \int \Delta\varepsilon(\vec{k}_1) E_s(\vec{k} - \vec{k}_1) d\vec{k}_1 \tag{S3}$$

The term $G(k) = \frac{k_0^2}{(k^2 - \varepsilon_0 k_0^2)}$ is the so-called Green function (more precisely, the spatial Fourier transform of the Green function) which defines the response of the system (described by the particular model) to a delta function excitation in space (equivalently, filtering in *k*-domain).

Considering a series expansion for the scattered field: $E_s = \sum_0^n E_s^n$, the implicit equation, eq. (S3), can be solved iteratively in the following way:

$$E_s^n(\vec{k}) = \frac{G(k)}{2\pi} \int [\Delta\varepsilon(k - k') E_s^{n-1}(k')] dk' \tag{S4}$$

where *n* represents the order of the scattering phenomena. The first iteration of the series, $E_s^1(\vec{r})$, is referred as the first-order Born approximation, usually known as Born approximation.

In case of the Schrödinger equation, the stationary field in spatial Fourier domain can be written as:

$$E_s(\vec{k}) = \frac{k_0^2}{k^2} \frac{1}{2\pi} \int \Delta\varepsilon(\vec{k}_1) E_s(\vec{k} - \vec{k}_1) d\vec{k}_1 \tag{S5}$$

We note that the sole difference between (S3) and (S5) is a new Green function, $G(k) = \frac{k_0^2}{k^2}$. Thus, eq. (S4) can also be used to calculate the scattered field, of different Born orders, for the Schrödinger equation, having in this case a different Green function.

Mathematically, the Schrödinger and Helmholtz equations represent two systems with slightly different Green functions, as evident from the scattering functions (S3) and (S5), with identical convolution integrals but different terms, $1/(k^2 - \varepsilon_0 k_0^2)$ in the Helmholtz case and $1/k^2$ in the Schrödinger case. The rest of the Hilbert transform remains identical. Physically, this means that the scattering processes results in different projections of the scattered waves into the free space propagation modes.

This small difference only affects the calculation of higher Born-approximation orders iteratively. Such a difference is not physically significant in our proposed approach, since the local directionality is warranted by the zeroing $\Delta\varepsilon(\vec{k})$ over half-plane, regardless of the scattering amplitude. The Helmholtz model allows describing the propagation of a planar wave guided mode in a 2D plane, being $k_0$ the wavenumber of the horizontal mode; for instance, the lowest order wave guided mode, where the fields propagate horizontally, and $k_0$ would be practically the wavenumber of a plane wave with that frequency, $k_0 = \omega/c$. For a somewhat thicker planar structure, it may be some other mode. Finally, in the extreme case of an infinitely thick structure, the field would propagate vertically. Then the wavenumber would become $k_0 = 0$, leading to the Schrodinger equation. Thus, the scattering described by Helmholtz model becomes equivalent to the Schrödinger equation for a vertically propagating mode. In the following sections, we study basic directionality fields in the form of sink to verify our theory beyond the paraxial approximation.

## 2.2. Sink directionality field with gain-loss media

Here, we perform FDTD simulations in an optical medium with an octagonally-patterned background potential. Figure 1(a) shows the spatial refractive distribution of the initial background profile, while the imaginary profile of the modified medium, generated by the directionality field $\vec{p}(\vec{r}) = -\vec{r}/|\vec{r}|$ in the form of a sink, is depicted in Fig. 1(b). A source placed within the modified medium injects electromagnetic waves with an electric field polarized perpendicularly to the plane. Figure 2 collectively shows the full-wave FDTD simulation results at various time frames, and for sources with different beam waist. Furthermore, the detailed time evolutions of the fields, also compared with the time evolution of the field inside the initially given medium, are provided in Supplemental Movie 3.

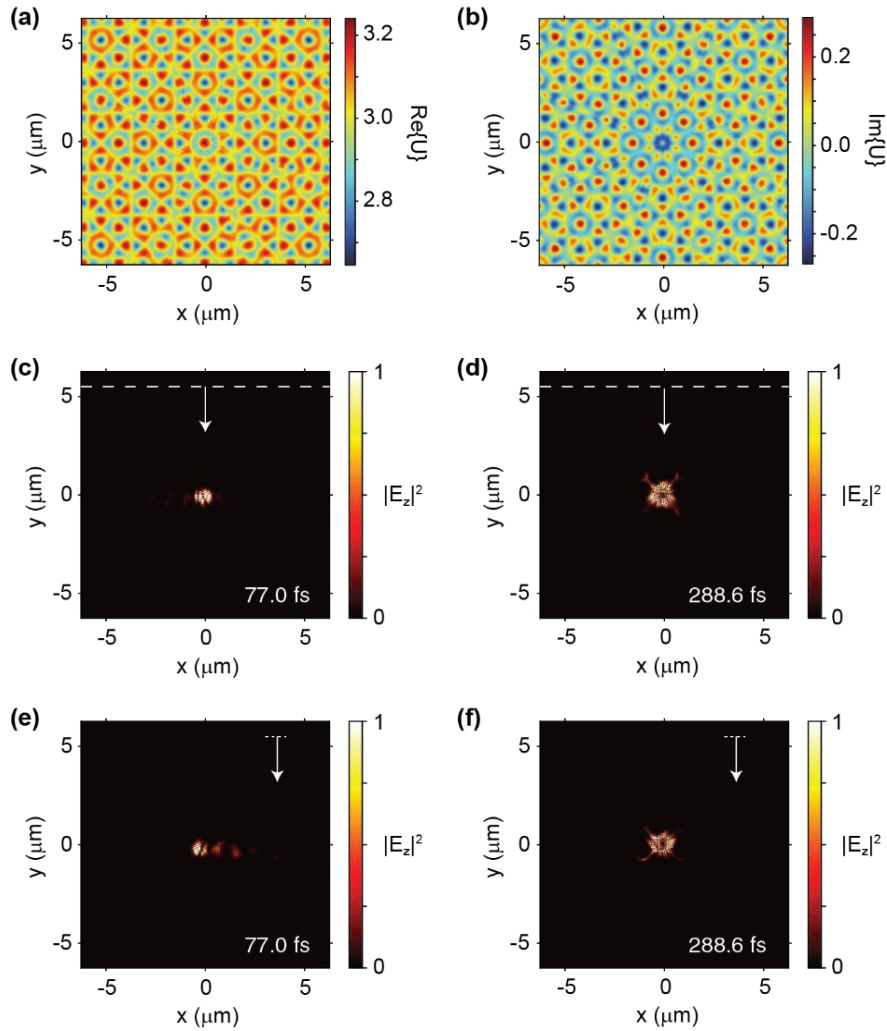

FIG. 1. (color online) Real (a) and imaginary (b) parts of the refractive index generated by the local Hilbert transforms in form of a sink. Instantaneous spatial intensity profiles for a line source illumination with a beam waist radius of 5.0 µm, at the time frames: (c) 77.0 fs and (d) 288.6 fs. Similar intensity profiles for a narrow-line source illumination, with a beam waist radius of 1.0 µm, at the time frames: (e) 77.0 fs and (f) 288.6 fs. Both sources inject a broadband pulse with central frequency of 599.6

THz (*i.e.* central wavelength of 500 nm) with a bandwidth of 133.6 THz. The dashed white lines indicate the positions of the sources, while the white arrows denote the source injection direction. The time evolutions of the intensity profiles are provided in Supplemental Movie 3. Numerical calculations were performed using Lumerical FDTD Solutions.

Note that for the case represented in Fig. 1, we observe the localization effect twice (see the Supplemental Movie 3). First, the incident pulse in form of plane wave propagating towards the sink, focuses at the center of the sink (see Figs. 1(c) and 1(e)). This is a transient behavior, and it evidences that the directionality field concentrates the incoming wave into the sink. The field behavior after this primary focusing strongly depends on the specific situation. In this particular case, the fields temporally dissipate from the sink position, however in a long-term scale, they again concentrate at the sink position, as clearly shown in Figs. 1(d) and 1(f)). We therefore show the twofold effect of field-management by the directionality field: transient and asymptotic. The asymptotic concentration of the probe field precisely corresponds to the (stationary) scattering theory considered in the main part of the paper.

### 2.3. Sink Directionality field with pure lossy media

An important advantage of the field directionality concept is that the complex locally transformed media can be constructed excluding gain materials and only considering loss materials. To achieve this goal, we simply add an "offset" to the imaginary part of the dielectric permittivity such that the gain areas are completely eliminated in the sample optical medium. Note that since the Fourier spectra of a constant offset function is a Dirac delta function positioned at the center of the wavevector domain, adding the offset does not distort the local spectra of the medium. We apply this method to the potential of Fig.1; the corresponding numerical calculations are given in Figs. 3(a) and 3(b) for sources with different waist radii. It follows from these figures, that the incoming wave is, as expected, directed towards the center, forming a localization around the center region. Important to note is that in this case no post-localization occurs, due to the absence of any gain material.

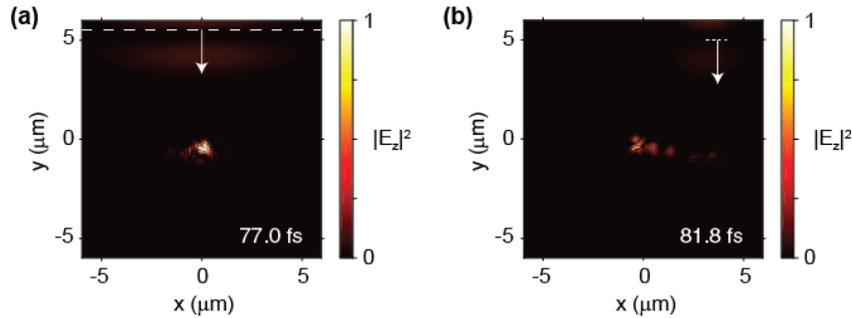

FIG. 2. (color online) Instantaneous spatial intensity profiles in a pure lossy medium: (a) for a line source illumination with a Gaussian beam waist radius of 5.0 µm and (b) for a narrow-line source illumination with a Gaussian beam waist radius of 1.0 µm at the time frames 77.0 fs and 81.8 fs, respectively. Temporal properties of the sources are the same as in Fig. 1. The dashed white lines indicate the positions of the sources, while the white arrows denote the source injection direction. Numerical calculations were performed using Lumerical FDTD Solutions.

The general conclusion is that the proposed theory of directionality fields constructed using local Hilbert transform can be applied not only to paraxial models like the Schrödinger equation, but also to more complicated wave propagation models, in particular to full Maxwell equations as simulated using the FDTD model. Such directionality field concept applied to the electromagnetic spectrum can lead to a plethora of optical applications ranging from efficient light detectors/absorbers, lasers with improved brightness and beam emission quality to optical data processors.